# Intrinsic Layer-Dependent Surface Energy and Exfoliation Energy of van der Waals Materials

Lin-Lin Wang[1,2*], Jiaqiang Yan[3], Yong Han[1], Claire C. Wang[4], Jian-Xiang Qiu[5], Su-Yang Xu[5], Adam Kaminski[1,2], Michael C. Tringides[1,2] and Paul C. Canfield[1,2]

[1]Ames National Laboratory, Ames, IA 50011, USA
[2]Department of Physics and Astronomy, Iowa State University, Ames, IA 50011, USA
[3]Oak Ridge National Laboratory, Oak Ridge, TN 37831, USA
[4]Ames High School, Ames, IA 50010, USA
[5]Department of Chemistry and Chemical Biology, Harvard University, Cambridge, MA 02138, USA

*llw@ameslab.gov

# Abstract

Stacking and twisting 2D van der Waals (vdW) layers have become versatile platforms to tune electron correlation. These platforms rely on exfoliating vdW materials down to a single and few vdW layers. We calculate the intrinsic layer-dependent surface and exfoliation energies of typical vdW materials such as, graphite, h-BN, black P, $MX_2$ (M=Mo and W, X=S, Se and Te), MX (M=Ga and In, X=S, Se and Te), $Bi_2Te_3$ and $MnBi_2Te_4$ using density functional theory. For exchange-correlation functionals with explicit vdW interaction, a single vdW layer always has the smallest surface energy, giving a surface energy reduction when compared to thicker vdW layers. However, the magnitude of this surface energy reduction quickly decreases with increasing number of atomic layers inside the single vdW layer for different vdW materials. Such atomic-layer-dependence in surface energy reduction helps explain the different effectiveness of exfoliation for different vdW materials down to a single vdW layer.

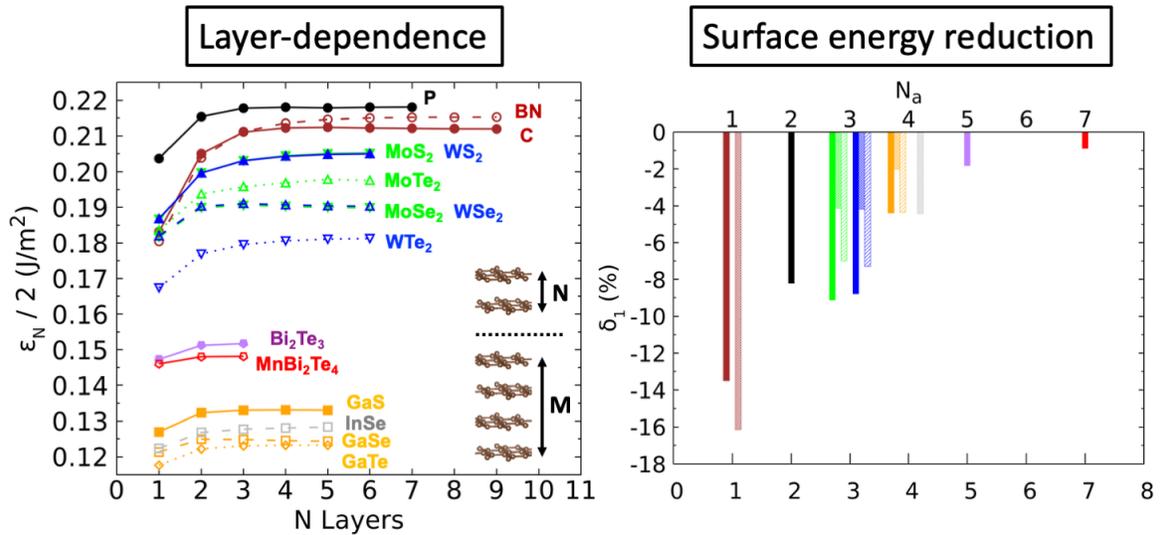

Van der Waals (vdW) interaction is ubiquitous in chemistry and materials science at various interfaces. As one of the basic chemical interactions, vdW is usually introduced in high school chemistry textbook[1]. More recently in physics, two-dimensional (2D) layered crystal structures bonded by vdW interaction such as graphite and transition metal dichalcogenides (TMDC) with their twisted stacking few layers have been found to be versatile systems that can be tuned for flat bands and non-trivial band topology to realize exotic phases of strong electron correlation. Graphitic materials have long been used as lubricants, electrodes and pencil cores. Its single layer form of graphene was first exfoliated via scotch tape method in 2004, resulting in an ideal 2D honeycomb lattice for C atoms[2]. Graphene is a quantum spin Hall (QSH) insulator[3] with a very small band gap at the $K$-point of the Brillouin zone (BZ) and is effectively a Dirac semimetal at finite temperature with high carrier mobility. Twisting bilayer graphene at a magic angle of ~1.1 degree can give moiré flat bands with strong electron correlation to realize superconductivity[4] and Mott insulator[5]. More recently 5-layer rhombohedral graphene[6] with flat bands have been found to host fractional quantum anomalous Hall effect (FQAHE). TMDC is another example of well-known vdW materials, which have been used as catalysts and lubricants in bulk form. But the 3-atomic-layer thick single vdW layer of TMDC, once exfoliated, can give interesting optoelectronic properties[7]. The twisted form of $MoTe_2$ has also been recently found to host FQAHE[8-10]. Similarly, vdW layered $Bi_2Te_3$ has long been used as a thermoelectric material before it was found as the prototype 3D topological insulator[11-14]. More recently, $MnBi_2Te_4$, a new member of this extended family of vdW materials, has been discovered as an intrinsic antiferromagnetic (AFM) topological insulator[15], with its few-layer forms being found to host axion and Chern insulators[16,17], as well as to realize quantum metric nonlinear Hall effect[18], a diode effect[19] and even axion quasiparticles[20]. All these novel layered stacking systems need the exfoliation of the corresponding vdW materials down to a single and few layers, which will also apply for future vdW materials in different lattice types[21].

However, despite the rapidly growing interest in these exfoliatable vdW materials, their intrinsic layer-dependent surface and exfoliation energies have not been well studied, in contrast to the well-known quantum size effect of thin metal films[22]. Experimentally, it can be challenging due to the impurities and defects prone for intercalation between vdW layers. Theoretically, a symmetric density functional theory[23, 24] (DFT) study of layer-dependent surface energies for vdW materials is also lacking despite the recent developments of vdW[25-27] exchange-correlation (XC) and meta-GGA[28-30] functionals. Previous DFT studies[31-33] have focused on the interlayer binding energy of vdW materials in bulk to benchmark different XC functionals with some consideration of bilayer interaction[33] and exfoliation of a single layer[31]. But when peeling off a vdW material with the scotch tape method, exfoliating a thin few layers instead of a single layer is likely to happen before reaching the single layer. In addition, twisting a few layers on top of another few layers instead of just bilayer twisting are also of increasing interest[34]. Thus, a systematic study of the layer-dependent surface energies for vdW materials is needed.

Here we calculate the layer-dependent surface energy of typical vdW materials of graphite, h-BN, black P, $MX_2$ (M=Mo and W, X=S, Se and Te), MX (M=Ga and In, X=S, Se and Te), $Bi_2Te_3$ and $MnBi_2Te_4$ with increasing number of atomic layers inside one vdW layer. Besides getting the ideal intrinsic surface energy at the bulk limit ($\gamma_\infty$), we also explore the trends of the layer-dependent surface energy ($\gamma_N$), where $N$ is the number of vdW layers. For the XC functionals with explicit vdW interaction, we find that the calculated $\gamma_N$ for a single layer at $N=1$ is always smaller than the thick layers before approaching the bulk limit at large $N$. Using graphite with PBE+D3 as an example, we decompose the vdW contributions to $\gamma_N$ to show that this behavior of surface energy reduction at very thin vdW layers is due to the less vdW interaction that need to be broken in the surface region of thin layers than the thick ones. Such surface energy reduction at very thin vdW layers helps explain the effectiveness of the scotch tape method, where the probability to create one surface of a very thin layer plus the other surface of a thick layer is higher than the probability to create two surfaces of thick layers because of the smaller energy cost of the former case from the surface energy reduction. We find that the surface energy reduction for $N=1$ is a generic behavior for vdW materials with different number of atomic layers ($N_a$) inside a single vdW layer, such as $N_a=1$ for graphite and h-BN, 2 for black P, 3 for $MX_2$, 4 for MX, 5 in $Bi_2Te_3$ and 7 in $MnBi_2Te_4$. But the magnitude of surface energy reduction quickly decreases with increasing $N_a$ from −16% for h-BN and −13% for graphite, −8% for black P, −4 to −9% for TMDC, −2 to −4% for MX, −2% for $Bi_2Te_3$ and only −1% for $MnBi_2Te_4$. This also helps explain the less effectiveness to peel off a single layer of vdW materials with thicker atomic layers, such as MnBi$_2$Te$_4$, than graphite. For the magnitude of the calculated $\gamma_\infty$, we found a range from 0.160 to 0.228 J/m$^2$ for graphite using different XC functionals with explicit vdW interaction, giving a better agreement with the recent experimental measurement[35] than LDA. With the more recent meta-GGA XC functional of r2SCAN+rVV10 benchmarked[32, 33] as the most accurate for vdW materials, the calculated $\gamma_\infty$ are 0.215 J/m$^2$ for h-BN, 0.218 J/m$^2$ for black P, 0.181 to 0.205 J/m$^2$ for TMDC, 0.123 to 0.133 J/m$^2$ for MX, 0.152 J/m$^2$ for Bi$_2$Te$_3$ and 0.148 J/m$^2$ for MnBi$_2$Te$_4$. These DFT-calculated intrinsic $\gamma_\infty$ provide references for future experimental measurements for these typical vdW materials.

DFT calculations have been performed with a plane-wave basis set and projector augmented wave method[36], as implemented in the Vienna Ab initio Simulation Package[37, 38] (VASP). A range of XC functionals of LDA[39, 40], PBE[41], PBEsol[42] and those with vdW interactions such as D2[43], D3[44, 45], vdWDF-BO[46], vdWDF-MK[27], and the more recent meta-GGA of SCAN+rVV10[29] and r2SCAN+rVV10[30] have been used. For each vdW material, the slabs with increasing $N$ vdW layers have been relaxed to calculate the layer-dependent surface energy with a large vacuum layer of at least 20 Å. For example with graphite, we have used a kinetic energy cutoff of 500 eV, $\Gamma$-centered Monkhorst-Pack[47] $k$-point mesh of (17×17×1) and a Gaussian smearing of 0.05 eV. The atomic positions and in-plane unit cell vectors for each slab are fully relaxed with the remaining absolute force on each atom being less than 1×10$^{-2}$ eV/Å. The surface energy ($\gamma_N$) as a function of $N$ vdW layers in the slab is defined as the energy cost to create two equivalent surface area ($A$) on both the top and bottom of slab with respect to the bulk energy ($\mu$) as

$$\gamma_N = \frac{1}{2A}(E_N - N\mu) \quad (1)$$

where the total energies of the fully relaxed slabs of $N$ layers ($E_N$) are calculated in DFT and used to fit Eq.(1) to find both $\mu$ and the $\gamma_N$ at bulk limit, i.e. $\gamma_\infty$ with large $N$. Then the $\mu$ can be used to calculate the layer-dependent $\gamma_N$ from Eq.(1) for slabs with small $N$.

Figure 1 plots the crystal structures of several typical vdW layered materials. Graphite, h-BN, majority of TMDC MX$_2$ (M=Mo, W and X=S, Se and Te) and MX (M=Ga and In, X=S, Se and Te) are all in space group (SG) 194 (*P6$_3$/mmc*). The two monolayers of C in graphite are each in the 2D honeycomb lattice and stacked in AB sequence or shifted in-plane by (1/3, 1/3), while the two monolayers of BN in h-BN are stacked in AA sequence with alternating element. In $MX_2$ and MX, one vdW layer unit consists of 3 and 4 atomic layers of X-M-X and X-M-M-X, respectively. The central M and M-M layer occupy the (1/3, 1/3) or (2/3, 2/3) site of the X sublattice, then these vdW layer units are stacked in AB sequence. The M and X sites share the same atomic column when viewed along the *c*-axis, but the X-M-X and X-M-M-X prisms are rotated by 180 degrees from the neighboring vdW layers. Unlike majority of $MX_2$ and MX, $WTe_2$ adopts an orthorhombic structure of SG-31 as shown in Fig.1(e), while InS and InTe do not crystalize in layered vdW structures and thus are not considered here. Black P has a base-centered orthorhombic structure of SG-64 as shown in Fig.1(c), where one vdW layer unit consists of two covalently bonded P atomic layers. $Bi_2Te_3$ and $MnBi_2Te_4$ are in SG-166 (*R–3m*) with the primitive unit cell consisting of one formula unit with one quintuple layer of Te-Bi-Te-Bi-Te and one septuple layer of Te-Bi-Te-Mn-Te-Bi-Te, respectively. These vdW layers are then stacked in the ABC sequence, giving the rhombohedral structure, which can also be viewed in a hexagonal conventional cell with three times of the primitive unit cell. The Mn in $MnBi_2Te_4$ has a magnetic moment of 5 $\mu_B$ with ferromagnetic alignment in the same septuple layer along the *c*-axis. Between the neighboring septuple layers, it has an A-type or up-down AFM configuration, so the $MnBi_2Te_4$ magnetic unit cell doubles its chemical unit cell. We chose these vdW layered structures in Fig.1 to study layer-dependent surface energy because of the increasing number of atomic layers $N_a$=1, 2, 3, 4, 5 and 7 inside the single vdW layer.

The calculated layer-dependent $\gamma_N$ of graphite are plotted in Fig.2 for different XC functionals. For graphite, it is well known that PBE does not bind between graphene layers because of lacking vdW interaction. In contrast, LDA binds between the layers despite of no explicit vdW interaction and gives a $\gamma_\infty$ of 0.066 J/m² at bulk limit in Fig.2(a), agreeing with the previous DFT study[48]. Before reaching the bulk limit, the layer-dependent $\gamma_N$ can have an even-odd behavior over *N* due to *k*-point mesh sampling the Dirac or band touching point at the *K* point (1/3, 1/3, 0) if the mesh is a multiplier of 3 (see Supporting Information Fig.S1 for *k*-point mesh convergence). For PBEsol, although there is binding between graphene layers and the relaxed bulk graphite structure is closer to the experiment than LDA, the calculated $\gamma_\infty$ is much smaller at 0.003 J/m², one order of magnitude smaller than LDA, which reflects the lack of vdW interaction in the regular XC functionals. With vdW interaction explicitly included in a range of XC functionals from the empirical D2, D3, vdW exchange functional vdWDF-BO, vdWDF-MK, and the more recent meta-GGA variants of SCAN+rVV10 and r2SCAN+rVV10, the calculated $\gamma_\infty$ are significantly larger than that of LDA. At bulk limit, the $\gamma_\infty$ is 0.189 J/m$^2$ for PBE+D2, 0.160 J/m$^2$ for PBE+D3, 0.226 J/m$^2$ for vdWDF-BO, 0.228 J/m$^2$ for vdWDF-MK, 0.185 J/m$^2$ for SCAN+rVV10, and 0.212 J/m$^2$ for r2SCAN+rVV10, agreeing with the previous DFT studies [31-33, 49] and also giving a range that is in a better agreement with the recent experiments than LDA. For graphite in experiment, some studies report $\gamma_\infty$, while most report binding energy, cleave energy or adhesion energy, which is twice the $\gamma_\infty$. After conversion, there is a range of experimentally reported $\gamma_\infty$ for graphite from 0.063 J/m$^2$ on the lower side[50] to 0.195 J/m$^2$ on the upper side[51], and others[35, 52-56] are in between. The most recent micro-force probing measurements[35] gave 0.164 J/m$^2$ and they also reported large effects from aging and air contamination depending on how the samples and interfaces were treated, which can cause $\gamma_\infty$ drop to 0.075 J/m$^2$. This reflects the difficulty in experiments to accurately measure the intrinsic $\gamma_\infty$ for vdW materials.

Between the two groups of XC functionals without and with vdW interaction in Fig.2, the most noticeable difference for $\gamma_N$ is the trend at the very thin slabs of $N$=1 and 2 vdW layers. For LDA and PBEsol without vdW interaction, $\gamma_N$ has an overall decreasing trend with increasing $N$, before approaching the bulk limit. The $\gamma_N$ for the very thin layers are larger than the bulk limit. However, for the XC functionals with explicit vdW interaction, the calculated $\gamma_N$ for $N$=1 and 2 are mostly smaller than those of the thicker slabs (except for $N$=2 for vdWDF-BO and vdWDF-MK), then $\gamma_N$ approaches the bulk limit at very large $N$. On first thought, this behavior of surface energy reduction for very thin slabs is counter intuitive, because when compared to the bulk with infinite number of vdW layers, thin slabs are less stable due to less vdW interaction among their small finite number of layers. Next, we will use PBE+D3 with reference to PBE to decompose the vdW contributions to $\gamma_N$ as a function of $N$ to show that the surface energy reduction at small $N$ is due to the less vdW interaction cost on the surface region to break for the thin layers than thick ones.

Starting from PBE without the D3 vdW interaction, because there is no binding between the graphene layers for PBE, we use the relaxed PBE+D3 slab structures to calculate the $E_N$ for PBE, and then PBE $\gamma_N$, which is plotted in Fig.3(a). The small negative value of PBE $\gamma_N$ shows that there is indeed no layer binding for graphite with PBE. But PBE $\gamma_N$ still decreases with increasing $N$, similar to those of LDA and PBEsol, i.e., without the surface energy reduction for $N$=1 and 2, in a distinct contrast to those $\gamma_N$ calculated from the XC functionals with vdW interaction.

With the added D3 for vdW interaction in PBE+D3, the new surface energy can be written similarly to Eq.(1) with the extra terms as the following

$$\gamma'_N = \frac{1}{2A}\left[(E_N + Nw_N) - N(\mu + \mu_{disp})\right] \qquad (2)$$

where $w_N$ is the vdW interaction per layer in the slab as in $w_N = w_{N6} + w_{N8}$ for the pairwise contributions of $r^{-6}$ and $r^{-8}$ exponential from dispersion interaction. The $\mu_{disp}$ is the D3 vdW correction in the bulk structure, acting as a constant shift of the bulk energy reference. For PBE+D3, $w_N$ can be explicitly calculated for slabs as a function of increasing $N$ and are plotted in Fig.3(b) with contributions from $w_{N6}$ and $w_{N8}$ terms. Thus, the additional contributions to $\gamma_N$ due to D3 vdW interaction can be written as the difference between Eq.(1) and (2) to give

$$\Delta\gamma_N = \frac{N}{2A}(w_N - \mu_{disp}) \qquad (3)$$

Note this additional contribution $\Delta\gamma_N$ due to D3 is proportional to $N$ as shown on the right side of Eq.(3). But when $N$ is large, this change should become a constant shift and independent of $N$ to approach the bulk limit. To better understand this behavior, we can write down the incremental difference between $N$ and $N-1$ slabs as the following,

$$\Delta\gamma_N - \Delta\gamma_{N-1} = \frac{1}{2A}\{[Nw_N - (N-1)w_{N-1}] - \mu_{disp}\} \qquad (4)$$

When $N$ is large, inserting an extra layer in the middle of a thick slab changes the vdW interaction per layer negligibly, so $w_N \approx w_{N-1}$, and the right side of Eq.(4) becomes $w_N - \mu_{disp}$, which approaches zero for the bulk limit because $w_\infty = \mu_{disp}$. Thus, the vdW contribution $\Delta\gamma_N$ to $\gamma_\infty$ amounts to a constant shift as shown in Fig.3(c) for large $N$.

Interestingly, as shown in Fig.3(b), $w_N$ has the largest change for the very thin slabs with much less negative value, because of reduced vdW interactions from the neighboring layers. For example, $N=1$ has no nearest neighbor and $N=2$ has no next nearest neighbor vdW layer interaction. Given the large changes in $w_N$ for the very thin slabs of $N=1$ and $N=2$, the vdW contribution to surface energy in Eq.(3) $\Delta\gamma_1 = (w_1 - w_\infty)$ for $N=1$ can be smaller than $\Delta\gamma_2 = 2(w_2 - w_\infty)$ for $N=2$, because of the factor of 2 and also $w_\infty$ for the overall bulk vdW interaction being a more negative value than $w_2$. Thus, for $\Delta\gamma_N$ as calculated in Eq.(3) and plotted in Fig.3(c), $\Delta\gamma_1 = 0.173$ J/m² is the smallest among all the thicknesses, followed by $\Delta\gamma_2$, and then $\Delta\gamma_N$ quickly increases and approaches the bulk limit with large $N$. When adding this $\Delta\gamma_N$ from D3 vdW interaction to PBE and plotting in Fig.3(d), the summed $\gamma_N$ agrees well with the $\gamma_N$ directly calculated with PBE+D3 and gives the behavior of surface energy reduction at small $N$, thus, explaining the very thin slabs having the smaller energy cost of $\gamma_N$ than the thick slabs. The deviation at large $N$ in Fig.3(d) is due to the small underestimation of D3 $w_\infty$ from $\mu_{disp}$, and then this small difference propagates linearly for large $N$ in Eq.(3). For accurate $\gamma_N$, the direct slab calculation should be used. The above decomposition and analysis of $\Delta\gamma_N$ contributions from D3 vdW interaction fully explain the surface energy reduction at small $N$. This surface energy reduction for the very thin layers of graphite is also reproduced for other vdW interactions as shown in Fig.2(c-h) for the vdW density functional of vdWDF-BO and vdWDF-MK, and also the more recent SCAN+rVV10 and r2SCAN+rVV10, thus, a generic feature for vdW interaction.

This surface energy reduction at very thin vdW layers does not mean that vdW materials will break into very thin layers by themselves, because the total surface area from the sum of thin layers is still much larger than the total surface area by breaking into thick layers, or in terms of total energy for the whole system, thick layers are still more stable than thin layers. But for creating a new surface area with scotch tape method, the smaller the $\gamma_1$ is for a single vdW layer, it means less energy cost and thus the probability to get a single vdW layer being peeled off the bulk vdW crystal is much higher. The case in LDA is the opposite by giving higher energy cost for thin layers, which means creating new surface area down to a single vdW layer is not energetically preferred when peeling off vdW materials. These are the ideal $\gamma_N$ for perfect surface structures, but real vdW materials tend to have defects and contain intercalated impurities and small molecules, which can change the magnitude of the measured surface energy. This is also the primary reason that it is difficult to measure the intrinsic ideal surface energy of vdW materials in experiment.

After establishing and explaining the surface energy reduction for very thin layers of graphite, next we study the $\gamma_N$ calculated in r2SCAN+rVV10 and their trends for other vdW materials with increasing number of atomic layers inside a single vdW layer. As plotted in Fig.4(a), the intrinsic $\gamma_\infty$ of 0.215 J/m² for BN is slightly larger than 0.212 J/m² for graphite. Although 0.218 J/m² for black P is even larger, its single vdW layer $\gamma_1$ is also significantly larger, making the magnitude of surface energy reduction for black P smaller than graphite and h-BN because of two atomic layers inside a single black P vdW layer.

As plotted in Fig.4(b), $\gamma_N$ of TMDC approach the bulk limit at smaller $N$ than graphite because of the thicker 3-atomic vdW layer. MoS$_2$ has a slightly larger $\gamma_\infty$ of 0.205 J/m² than 0.198 J/m² for MoTe$_2$ and then followed by 0.190 J/m² for MoSe$_2$. These values are only slightly smaller than 0.212 J/m² for graphite. The calculated $\gamma_\infty$ for MoS$_2$ in r2SCAN+rVV10 agrees with those calculated with DFT+D2 and vdW-DF from a previous study[57], and is also close to the recent experimental measurement[35] of 0.241 J/m². Replacing Mo with W, the $\gamma_N$ of WS$_2$ and WSe$_2$ is almost the same as MoS$_2$ and MoSe$_2$, respectively, showing the dominant contribution from the boundary atoms in the vdW layer with the same crystal structure. The $\gamma_N$ of WTe$_2$ is smaller than the other TMDC because it adopts a different crystal structure in SG-31. But the shape of $\gamma_N$ vs $N$ for WTe$_2$ is similar to that of MoTe$_2$ with nearly a constant downshift of 0.016 J/m². Importantly, the surface energy reduction, i.e., the $\gamma_1$ being smaller than thick layers still remains for all the TMDC, regardless of two different crystal structures, explaining the effectiveness for exfoliating single TMDC vdW layer with the scotch tape method. For MX (M=Ga and In, X=S, Se and Te) with $N_a$=4 atomic-layered vdW structure, their $\gamma_N$ vs $N$ in Fig.4(c) have similar trends to those of TMDC in Fig.4(b), but the magnitude of $\gamma_N$ decreases significantly to around 0.130 J/m² and smaller.

For Bi$_2$Te$_3$ and MnBi$_2$Te$_4$ with $N_a$=5 and 7 atomic-layered vdW structures, respectively, their $\gamma_N$ in Fig.4(d) mostly already converge to the bulk limit at just $N$=2 because of the thick single vdW layer. The bulk limit $\gamma_\infty$ of 0.152 J/m² for Bi$_2$Te$_3$ is slightly larger than 0.148 J/m² for MnBi$_2$Te$_4$. Both values are significantly smaller than those of TMDC and graphite. The single vdW layer $\gamma_1$ still is smaller than the thick layers for both compounds showing a generic behavior of surface energy reduction at the single layer for different vdW materials. But their surface energy reductions are very small at only 0.001 J/m$^2$ for MnBi$_2$Te$_4$ and 0.003 J/m$^2$ for Bi$_2$Te$_3$. To further quantify the surface energy reduction, we use the percentage change of $\gamma_N$ between the single vdW layer and bulk limit

$$\delta_1 = (\gamma_1 - \gamma_\infty)/\gamma_\infty \tag{5}$$

as plotted in Fig.4(e) for different vdW materials with respect to $N_a$, i.e., the atomic layer thickness of single vdW layers. There is almost a linear relation between $\delta_1$ and $N_a$, −16% for h-BN and −13% for graphite ($N_a$=1), −8% for black P ($N_a$=2), −4 to −9% for TMDC ($N_a$=3), −2 to −4% for MX ($N_a$=4), −2% for Bi$_2$Te$_3$ ($N_a$=5), and only −1% for MnBi$_2$Te$_4$ ($N_a$=7). Such a scaling behavior can be understood from Eq.(2), because when the single vdW layer becomes thicker from 1 to 7 atomic layers, the vdW interaction as in $w_N$ and $\mu_{disp}$ becomes smaller in proportion to all the other chemical bonding, and vdW interaction also decays in $r^{-6}$ and $r^{-8}$ exponentials. Or in other words, the vdW interaction among 7 atomic-layered vdW slabs is mostly from the nearest neighbor vdW layers.

Lastly, we will show that the exfoliation energy ($\mathcal{E}_N$) (cleave energy or adhesion energy) is also layer-dependent and is equivalent to twice of $\gamma_N$. For the exfoliation of $N$ vdW layers from a thick ($N$+$M$) slab as depicted in the inset of Fig.4(f) ($M \gg N$), the layer-dependent $\mathcal{E}_N$ can be calculated directly from the total energies of the three slabs as

$$\mathcal{E}_N = \frac{1}{A}(E_N + E_M - E_{N+M}) \tag{6}$$

The slab total energy $E_N$ is also related to the $\gamma_N$ and $\mu$ via Eq.(1) as $E_N = N\mu + 2A\gamma_N$. After substituting these terms into Eq.(6), we get

$$\mathcal{E}_N = 2\gamma_N + 2(\gamma_M - \gamma_{N+M}) \tag{7}$$

When $M$ is sufficiently large approaching to the bulk limit, $\gamma_M \approx \gamma_{N+M} \approx \gamma_\infty$, the layer-dependent exfoliation energy $\mathcal{E}_N$ becomes

$$\mathcal{E}_N \approx 2\gamma_N \tag{8}$$

To verify this numerically, we have calculated the $\mathcal{E}_N$ directly from the slab total energies using Eq.(6) with r2SCAN+rVV10 for the different vdW materials and plotted $\mathcal{E}_N/2$ in Fig.4(f). Note here $N$ only goes up to half of the available slab thickness to satisfy the bulk limit for $M$. Comparing the layer-dependent $\mathcal{E}_N/2$ in Fig.4(f) to those $\gamma_N$ in Fig.4(a) for graphite, h-BN and black P, Fig.4(b) for TMDC, Fig.4(c) for MX, and Fig.4(d) for Bi$_2$Te$_3$ and MnBi$_2$Te$_4$, they all agree very well at both small and large $N$, validating Eq.(8). Thus, the layer-dependent $\mathcal{E}_N$ behaves the same as $\gamma_N$ for vdW materials with a reduction at very thin layers mostly of $N$=1 and 2. Such reduction becomes smaller with increasing number of atomic layers $N_a$ in the single vdW layer as shown in Fig.4(e). This trend helps explain the different effectiveness of the scotch tape method and means that peeling off a single MnBi$_2$Te$_4$ layer of 7-atomic layers is more time-consuming than the 1-atomic layer thick graphene, because the probability to break into two thick vs a single plus a thick MnBi$_2$Te$_4$ layer are almost the same due to the very small surface energy difference.

We have also investigated the possible correlation between the equilibrium vdW interlayer distance ($d_{vdW}$) averaged over the slab and $\gamma_\infty$ as plotted in Supporting Information Fig.S2. There is a correlation between increasing $d_{vdW}$ and decreasing $\gamma_\infty$ mostly for both the same crystal structure and same boundary atomic types, however, across different crystal structures or different boundary atomic types, there is no overall correlation. Future systematic endeavors are needed to address the effects of intrinsic defects like vacancies and extrinsic defects like small molecules intercalated inside the vdW gaps on the surface energy and exfoliation energy of various vdW materials as functions of chemical potentials according to the experimental growth and exfoliation conditions.

In conclusion, we have calculated and explored the trend of layer-dependent surface energy ($\gamma_N$) and exfoliation energy ($\mathcal{E}_N$) in DFT for typical vdW materials of graphite, h-BN, black P, MX$_2$ (M=Mo and W, X=S, Se and Te), MX (M=Ga and In, X=S, Se and Te), Bi$_2$Te$_3$ and MnBi$_2$Te$_4$ with increasing number of atomic layers inside one vdW layer. For the XC functionals with explicit vdW interaction, we find a generic behavior of surface energy reduction for very thin vdW layers of $N$=1 and sometime $N$=2, where their calculated $\gamma_N$ are smaller than the thick layers before approaching the bulk limit at large $N$. Using PBE+D3 for graphite as an example, we decompose the vdW contributions to $\gamma_N$ to show that this behavior of surface energy reduction for very thin vdW layers is due to the less vdW interaction needs to be broken in the surface region for the thin slabs than thick ones. The surface energy reduction helps explain the effectiveness of scotch tape method to peel thin layers from vdW materials, where the probability to create one surface of very thin layer plus the other surface of a thick layer is higher than the probability to create two surfaces of thick layers, because the energy cost of the former is smaller than the latter. However, the magnitude of surface energy reduction decreases quickly from −16% for h-BN and −13% for graphite, −8% for black P, −4 to −9% for TMDC, −2 to −4% for MX, −2% for $Bi_2Te_3$ and only −1% for $MnBi_2Te_4$, when the number of atomic layers inside a single vdW layer increases from 1, to 2, 3, 4, 5 and 7, respectively. Such difference also helps explain the less effectiveness to peel off a single vdW layer from $MnBi_2Te_4$ than graphite. For the calculated $\gamma_\infty$, or half of $\mathcal{E}_\infty$, we find a range of 0.160 to 0.228 $J/m^2$ for graphite using different XC functionals with explicit vdW interaction, giving a better agreement with the recent experimental measurement[35] than LDA. With the more recent meta-GGA XC functional of r2SCAN+rVV10, the calculated $\gamma_\infty$ is 0.215 $J/m^2$ for h-BN, 0.218 $J/m^2$ for black P, 0.181 to 0.205 $J/m^2$ for TMDC, 0.123 to 0.133 $J/m^2$ for MX, 0.152 $J/m^2$ for $Bi_2Te_3$ and 0.148 $J/m^2$ for $MnBi_2Te_4$. These DFT-calculated intrinsic $\gamma_\infty$ provide references for future experiments to measure the surface energies of these vdW materials.

**Data Availability**: The data that support the findings of this study are openly available[58].

**Supporting Information**: K-point mesh convergence for layer-dependent surface energies of graphite in Fig.S1; Equilibrium vdW interlayer distances and surface energies in Fig.S2

**Author Contributions**: L.-L.W. and J.Q.Y. conceived and designed the work with inputs from P.C.C., A.K., M.C.T, S.-Y.X, J.-X.Q. and Y.H. L.-L.W. performed the ab initio calculations and data analysis with help from Y. H. and C.C.W. All the authors contributed to the discussion of the results, writing and review of the final manuscript.

**Competing Interests**: The authors declare no competing interests.

# Acknowledgements

This work was supported by the U.S. Department of Energy Office of Science, Office of Basic Energy Sciences through the Ames National Laboratory. The Ames

National Laboratory is operated for the U.S. Department of Energy by Iowa State University under Contract No. DE-AC02-07CH11358. J.Q.Y. was supported by the US Department of Energy, offce of Science, Basic Energy Sciences, Materials Sciences and Engineering Division. J.-X.Q. and S.-Y.X. were supported by the Center for the Advancement of Topological Semimetals, an Energy Frontier Research Center funded by the U.S. Department of Energy Office of Science, Office of Basic Energy Sciences through the Ames National Laboratory under its Contract No. DE-AC02-07CH11358. Some of this research used resources of the National Energy Research Scientific Computing Center (NERSC), a DOE Office of Science User Facility.

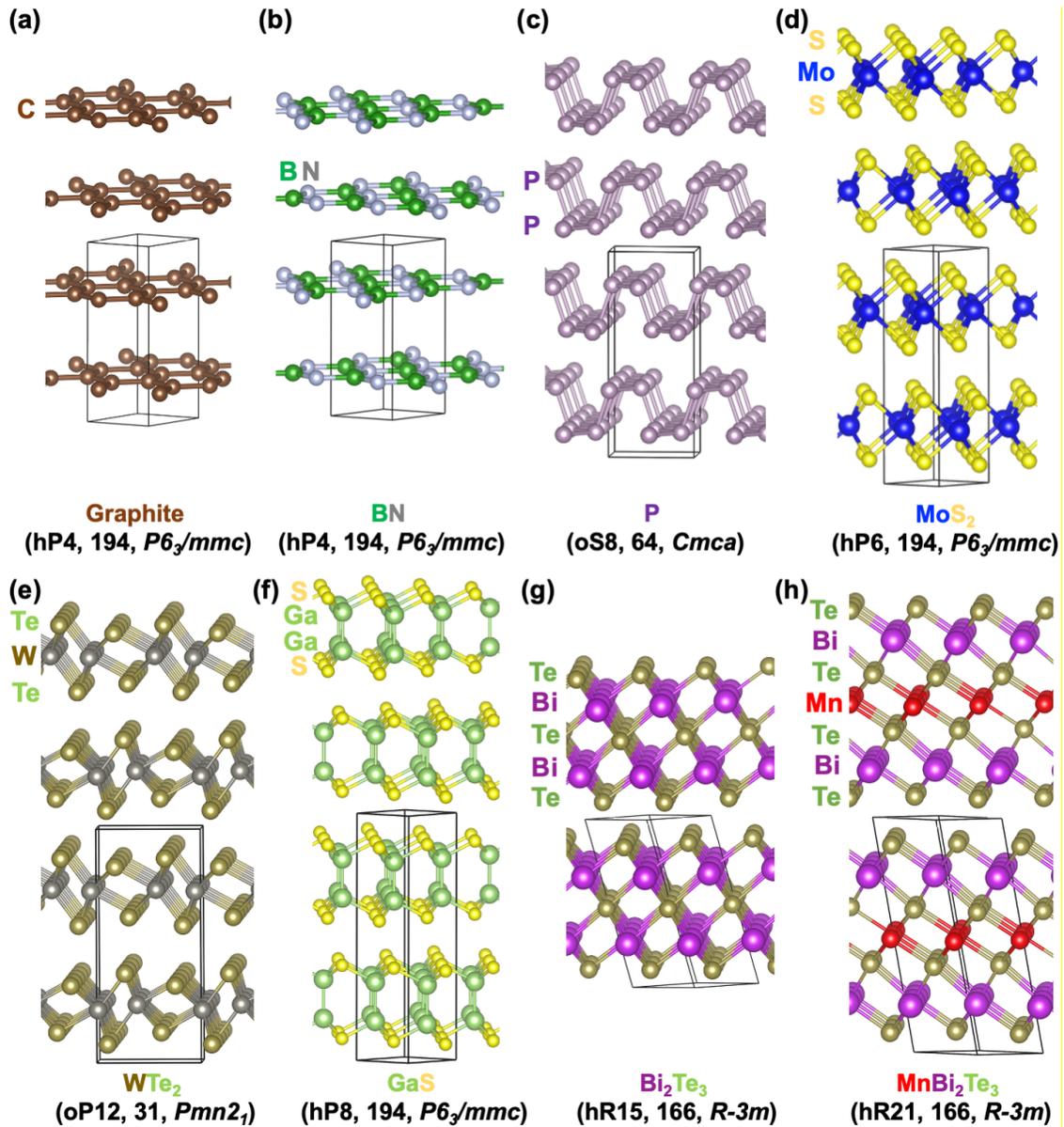

Figure 1. Typical van der Waals (vdW) layered crystal structures with increasing number of atomic layers in a single vdW layer. (a) Graphite, (b) h-BN, (c) black P, (d) $MX_2$ (M=Mo and W, X=S, Se and Te), (e) $WTe_2$, (f) MX (M=Ga and In, X=S, Se and Te), (g) $Bi_2Te_3$ and (h) $MnBi_2Te_4$. The atomic species are labeled in the corresponding colors to show the increasing number of atomic layers 1, 2, 3, 4, 5 and 7 in the different vdW layered structures. The Pearson notation, space group number and their labels are listed in the parenthesis with the unit cells highlighted.

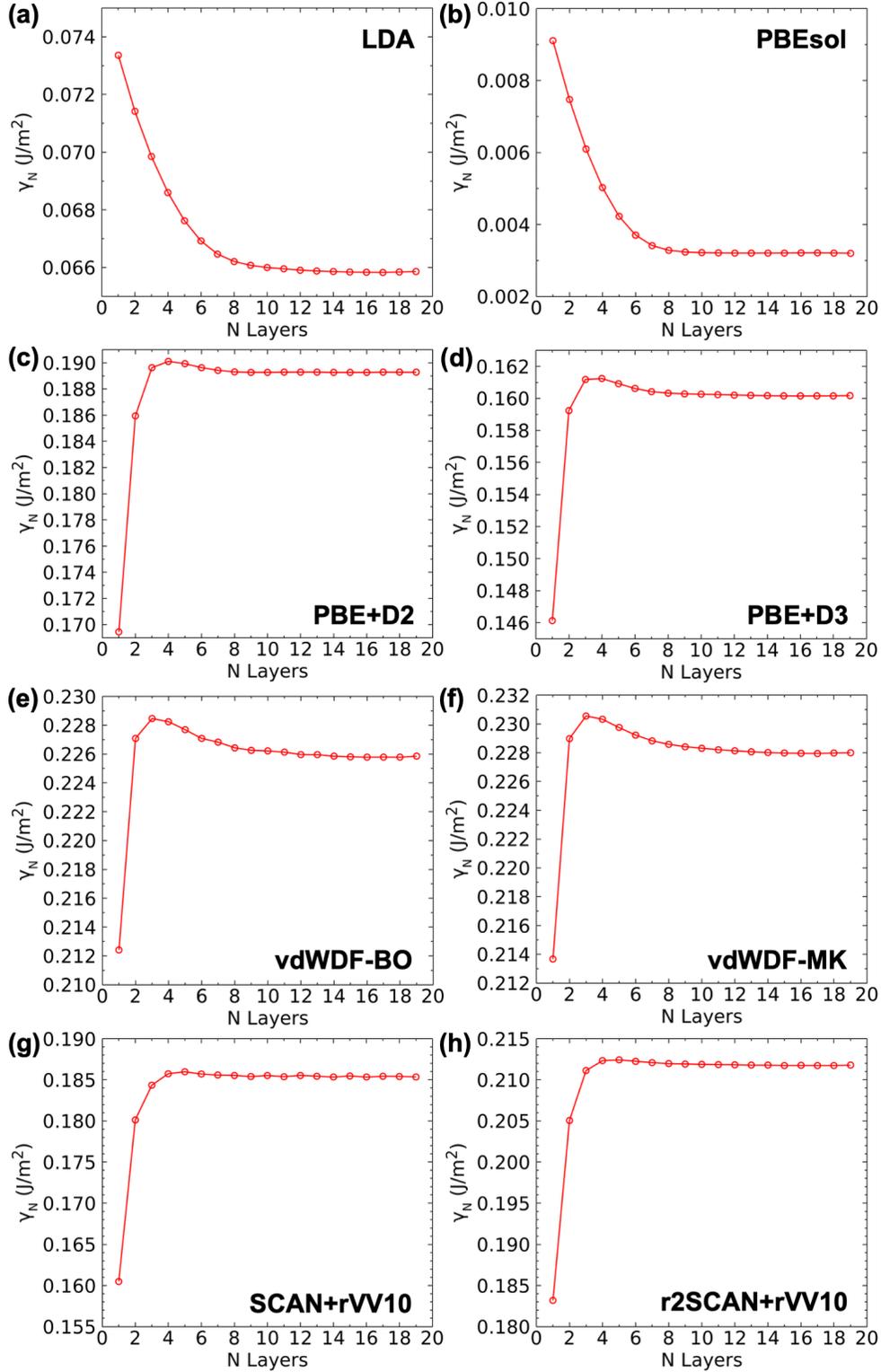

Figure 2. Layer-dependent surface energy ($\gamma_N$) of graphite as function of $N$ layers from 1 to 19 calculated in different exchange-correlation functionals. (a) LDA, (b) PBEsol, (c) PBE+D2, (d) PBE+D3, (e) vdWDF-BO, (f) vdWDF-MK, (g) SCAN+rVV10 and (h) r2SCAN+rVV10.

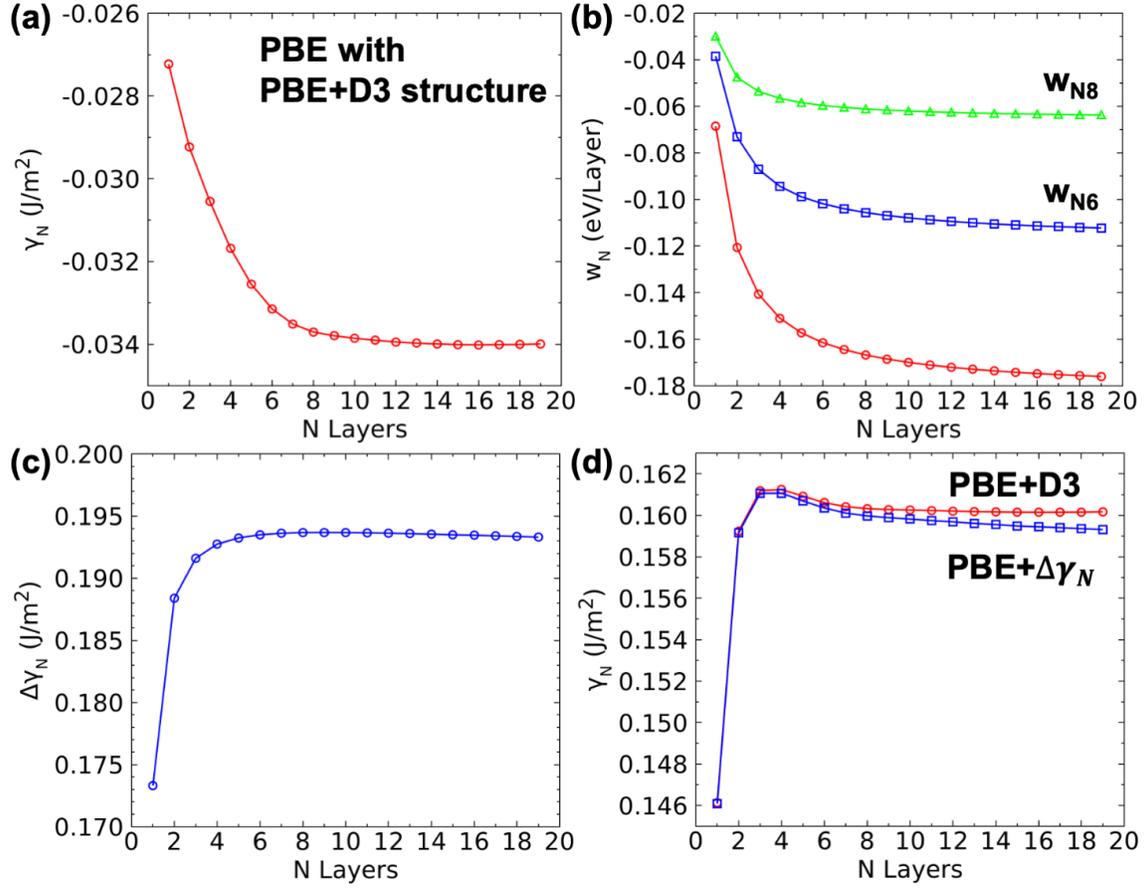

Figure 3. Decomposition of layer-dependent surface energy ($\gamma_N$) of graphite in PBE+D3. (a) $\gamma_N$ calculated in PBE with the relaxed structures in PBE+D3 as function of $N$ layers from 1 to 19. (b) D3 vdW interaction $w_N$ decomposed into $w_{N6}$ and $w_{N8}$ terms for $r^{-6}$ and $r^{-8}$ exponentials, respectively. (c) Surface energy difference ($\Delta\gamma_N$) in Eq.(3) due to D3 vdW interaction as function of $N$. (d) Comparison between the $\gamma_N$ calculated directly with PBE+D3 and the sum of PBE and the contribution due to D3.

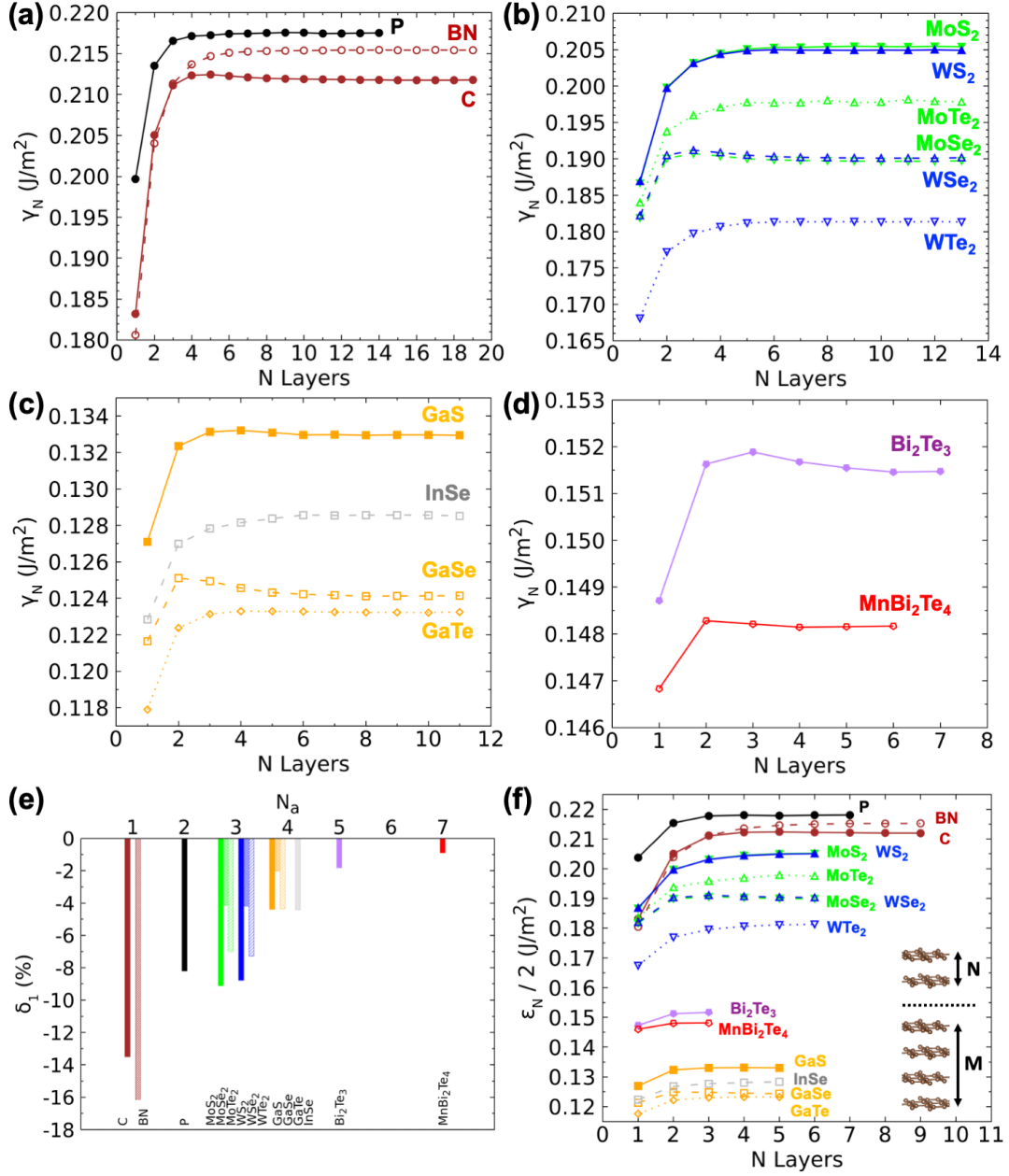

Figure 4. Layer-dependent surface energy ($\gamma_N$) and exfoliation energy ($\mathcal{E}_N$) of vdW materials as function of $N$ layers calculated in r2SCAN+rVV10. (a) $\gamma_N$ for graphite (C), h-BN and black P. (b) $\gamma_N$ for transition metal dichalcogenides $MX_2$ (M=Mo and W, X=S, Se and Te). (c) $\gamma_N$ for MX (M=Ga and In, X=S, Se and Te). (d) $\gamma_N$ for $Bi_2Te_3$ and $MnBi_2Te_4$. (e) Surface energy reduction ($\delta_1 = (\gamma_1 - \gamma_\infty)/\gamma_\infty$) defined as the relative change in percentage for a single vdW layer vs the converged bulk limit with increasing number of atomic layers ($N_a$=1, 2, 3, 4, 5 and 7) inside a single vdW layer for the different vdW materials as labeled. (f) Layer-dependent exfoliation energy $\mathcal{E}_N/2$ of vdW materials calculated via the exfoliation of $N$ vdW layers from a ($N+M$) slab (inset).

*Supporting Information*

# Intrinsic Layer-Dependent Surface Energy and Exfoliation Energy of van der Waals Materials

Lin-Lin Wang[1,2*], Jiaqiang Yan[3], Yong Han[1], Claire C. Wang[4], Jian-Xiang Qiu[5], Su-Yang Xu[5], Adam Kaminski[1,2], Michael C. Tringides[1,2] and Paul C. Canfield[1,2]

[1]Ames National Laboratory, Ames, IA 50011, USA
[2]Department of Physics and Astronomy, Iowa State University, Ames, IA 50011, USA
[3]Oak Ridge National Laboratory, Oak Ridge, TN 37831, USA
[4]Ames High School, Ames, IA 50010, USA
[5]Department of Chemistry and Chemical Biology, Harvard University, Cambridge, MA 02138, USA

*llw@ameslab.gov

# K-point mesh convergence for layer-dependent surface energies of graphite

The layer-dependent even-odd behavior in the calculated surface energy ($\gamma_N$) of graphite for (15×15×1) k-point mesh is due to the inclusion of the high-symmetry Dirac point at (1/3, 1/3, 0). Such even-odd behavior is reduced but remains at a much dense k-point mesh of (42×42×1) because of multiply of 3. In contrast, with a prime number k-point mesh to avoid the high-symmetry k-points in BZ integration for total energy, the calculated $\gamma_N$ does not show the even-odd behavior and is converged at (17×17×1) when compared to (41×41×1) k-point mesh.

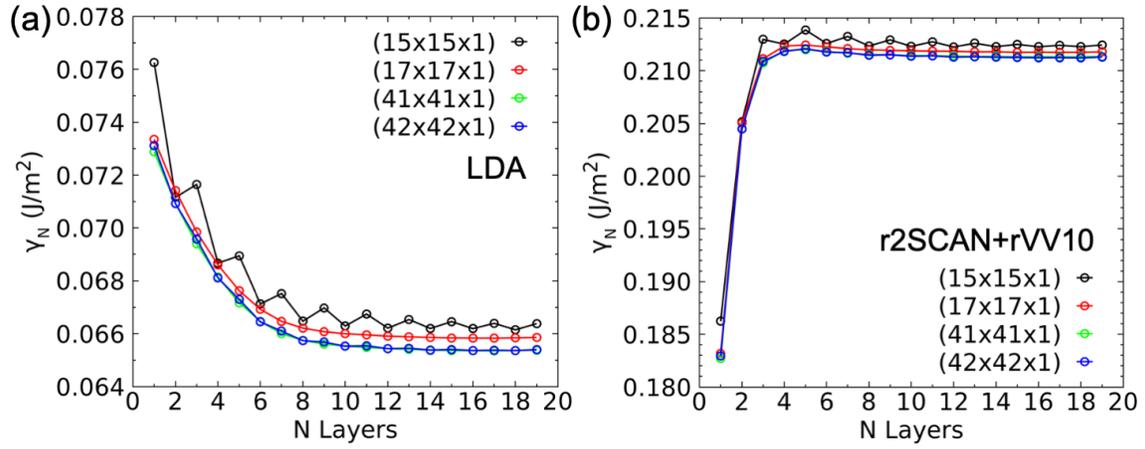

Figure S1. K-point mesh convergence for layer-dependent surface energy ($\gamma_N$) of graphite as function of $N$ layers. (a) LDA and (b) r2SCAN+rVV10.

# Equilibrium vdW interlayer distances and surface energies

We have investigated the correlation between the equilibrium vdW interlayer distance ($d_{vdW}$) averaged over the slab and the surface energy ($\gamma_\infty$) or exfoliation energy at bulk limit as plotted in Fig.S2. There is a correlation between increasing $d_{vdW}$ and decreasing $\gamma_\infty$ mostly when the vdW materials have both the same crystal structure and same boundary atomic types, as shown by the solid lines, such as the pairs of $MoS_2$-$WS_2$, $WTe_2$-$MoTe_2$ and InSe-GaSe. This is also true for $MoTe_2$-$WTe_2$ in the orthorhombic structure. This is even true for the case of BN-C with similar vdW/covalent radius of B, C and N, as well as $Bi_2Te_3$-$MnBi_2Te_4$ with the same outmost two boundary atomic layers being Bi-Te. However, overall, first with the same crystal structure but across different boundary atomic types, there is no such strong correlation as shown by the dashed lines for $MX_2$ and MX (X=S, Se and Te). For example, although $WTe_2$-$MoTe_2$ have larger $d_{vdW}$ than both $MoS_2$-$WS_2$ and $MoSe_2$-$WSe_2$, their $\gamma_\infty$ are in between these two later pairs. Secondly, across different crystal structures, although $MoS_2$ and GaS structures have the similar range of $d_{vdW}$ due to the similar vdW/covalent radius, the $\gamma_\infty$ changes a lot from 0.200 J/m$^2$ for $MoS_2$ structures to 0.130 J/m$^2$ for GaS-type vdW materials.

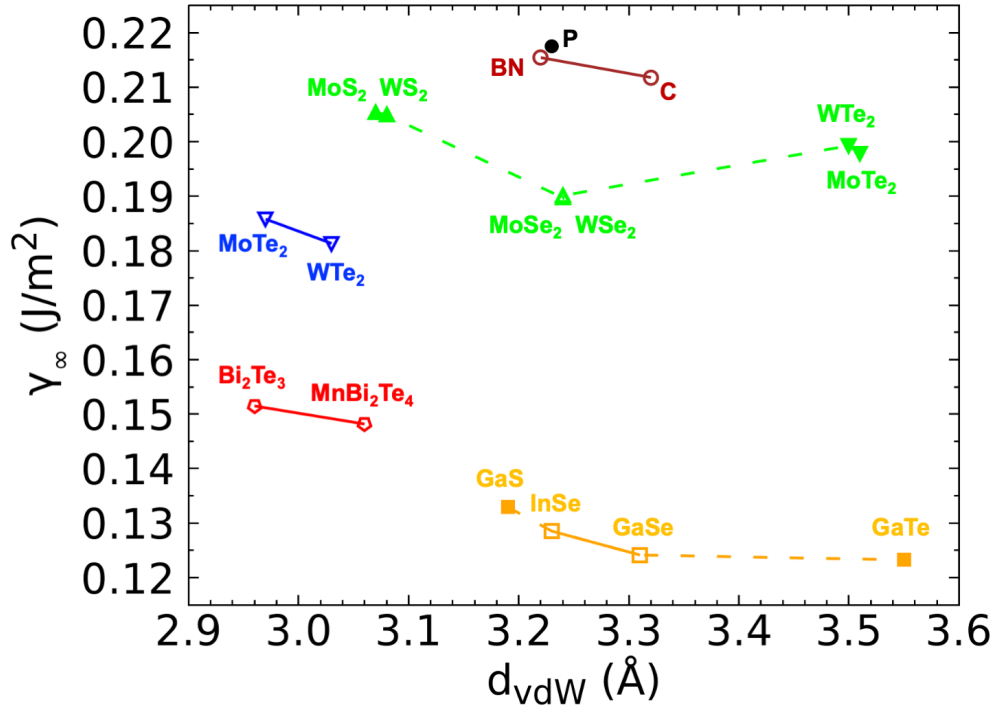

Figure S2. The equilibrium vdW interlayer distance ($d_{vdW}$) averaged over the slab vs the surface energy ($\gamma_\infty$) at bulk limit for different vdW materials.